# Prediction of Model Generalizability for Unseen Data: Methodology and Case Study in Brain Metastases Detection in T1-Weighted Contrast-Enhanced 3D MRI

Engin Dikici[1,*], Xuan Nguyen[1], Noah Takacs[1], Luciano M. Prevedello[1]

[1] The Ohio State University, College of Medicine, Department of Radiology, Columbus, OH, 43210, USA

[*] Corresponding author: Engin Dikici (engin.dikici@osumc.edu)

## Abstract

*Background and purpose:* A medical AI system's generalizability describes the continuity of its performance acquired from varying geographic, historical, and methodologic settings. Previous literature on this topic has mostly focused on "how" to achieve high generalizability (e.g., via larger datasets, transfer learning, data augmentation, model regularization schemes), with limited success. Instead, we aim to understand "when" the generalizability is achieved: Our study presents a medical AI system that could estimate its generalizability status for unseen data on-the-fly.

*Materials and methods:* We introduce a latent space mapping (LSM) approach utilizing Fréchet distance loss to force the underlying training data distribution into a multivariate normal distribution. During the deployment, a given test data's LSM distribution is processed to detect its deviation from the forced distribution; hence, the AI system could predict its generalizability status for any previously unseen data set. If low model generalizability is detected, then the user is informed by a warning message integrated into a sample deployment workflow. While the approach is applicable for most classification deep neural networks (DNNs), we demonstrate its application to a brain metastases (BM) detector for T1-weighted contrast-enhanced (T1c) 3D MRI. The BM detection model was trained using 175 T1c studies acquired internally (from the authors' institution) and tested using (1) 42 internally acquired exams and (2) 72 externally acquired exams from the publicly distributed Brain Mets dataset provided by the Stanford University School of Medicine. Generalizability scores, false positive (FP) rates, and sensitivities of the BM detector were computed for the test datasets.

*Results and conclusion:* The model predicted its generalizability to be low for 31% of the testing data (i.e., two of the internally and 33 of the externally acquired exams), where it produced (1) ~13.5 false positives (FPs) at 76.1% BM detection sensitivity for the low and (2) ~10.5 FPs at 89.2% BM detection sensitivity for the high generalizability groups respectively. These results suggest that the proposed formulation enables a model to predict its generalizability for unseen data.

## 1. Introduction

Artificial intelligence (AI) has been utilized immensely in medical applications over the last few decades [1], and the advent of deep neural networks (DNNs) has further expanded its adoption [2]. Recently developed AI applications (e.g., in diagnostics [3]–[5], prognostics [6]–[8], and treatment response prediction [9]–[11]) represent a giant leap forward from their counterparts only a few years ago; however, their widespread adoption is still restricted due to their limited generalizability [12]–[14]. Generalizability of an AI system is a broad concept describing the continuity of its performance when the data is coming from varying (1) geographic (e.g., institutions), (2) historical (e.g., timeframes), and (3) methodologic (e.g., acquisition parameters) settings [15]. Accordingly, limitations in generalizability manifest as poorer AI performance over time or when the deployment occurs across institutions with heterogeneous populations and imaging protocols [16].

In [17], generalizability in clinical research was presented as a hard-to-achieve goal, even as a myth, due to substantial context differences between institutions (caused by site-dependent items such as cohorts of patients and acquisition tools). The authors suggested that instead of seeking broadly generalizable tools, the research community should prioritize understanding how AI systems work and when and why they fail. Nonetheless, there is optimism about addressing an AI system's generalizability target during its development stage; [18] presented a weak correlation between the generalizability and a DNN's complexity (described using metrics such as its network size, norms, and sharpness). In their study, Eche et al. [16] identified the major causes of reduced generalizability in medical imaging systems as (1) overfitting (i.e., the AI-model learns unnecessary residual variations; information only applicable to training data [19]) and (2) underspecification (i.e., AI-model fails to learn the complete underlying statistical presentation of the data [20]). They argued that stress tests to evaluate a system's performance on shifted (i.e., generated via modifying data resolution or simulating different reconstruction kernels) or stratified datasets could enable the selection of models with reduced underspecification, where the solution was demonstrated on an AI model detecting hepatic steatosis in Computed Tomography (CT) datasets. The overfitting problem and its implications on AI-based radiology applications were further investigated in [21]. The study suggested that data augmentation [22], transfer learning [23], and model regularization [24] methods may alleviate the issue, whereas external validations are still necessary before a system's incorporation into clinical use to ensure its generalizability. In [14], the data heterogeneity between populations (i.e., clusters) was presented as a cause of reduced generalizability. It utilized electronic health records (EHR) to measure cluster heterogeneities and adopted internal-external cross-validation [25] to produce a multi-site model predicting the risk of atrial fibrillation from clinical data. Thus, the approach is only applicable when (1) the participant data (or EHR) is available and (2) the sites are willing to share, train, and update their models synchronously.

From a statistical perspective, reduced generalizability arises due to a mismatch between the probabilistic distributions of training and testing datasets: a phenomenon referred to as the domain shift [26]. As DNNs are data-driven models, large and representative [27] training datasets could mitigate the aforementioned causes of domain shift [28]. While there are public medical image databases [29] with many thousands of images (e.g., [30], [31]), (1) the number of datasets focusing on specific modalities and medical conditions is limited [32], and (2) widely used DNNs (e.g., [33], [34]) have many millions of trainable parameters, making generalizability unattainable in many clinical scenarios [35].

Unless (1) massive medical datasets with sufficiently high representation across historical, geographical, and methodical domains are built or (2) a wide range of institutions enter into agreements to train their medical AI models in harmony for the foreseeable future, the generalizability problem is here to stay. Adopting the skepticism raised by [17], we are not proposing yet another approach that may provide limited or conditional generalizability. Instead, our motivation is to introduce an AI model that is capable of quantifying its generalizability status for unseen data on-the-fly, enabling a real-time warning when the deployed model's generalizability is expected be low for a given case. Hence, we aim to understand "when" an AI model is generalizable rather than attempting to solve the broad problem of "how" to attain it, as previous literature on the topic has done. Our study introduces an AI-model formulation enabling the detection of reduced model generalizability for the given exam (i.e., temporal, on-the-fly), which may be occurring due to underspecification, overfitting, or other issue(s). It presents (1) a novel latent space mapping (LSM) approach to force training data into a known probabilistic distribution during model training, (2) the detection of reduced generalizability by processing test data LSMs, and (3) a sample integration of the reduced generalizability flag into an existing radiology workflow (see Fig. 1).

This report first provides a brief description of the AI application used for the validation study, a brain metastases (BM) detector for T1-weighted contrast-enhanced 3D MRI (T1c). Next, it describes the technical contribution and presents the merit of the approach via a case study in which the AI model is trained using data collected from our institution (175 T1c studies), then tested on **(1)** data collected internally (42 T1c studies) to confirm satisfactory model performance on new same-institution data, and **(2)** data collected from an external source (72 T1c studies from Stanford University's public BM dataset [36]) to assess model performance that is expected to be lower due to geographic and methodological differences in the imaging data. In this demonstration, the AI model computes and displays a metric indicating its generalizability status on a case-by-case basis during predictions. The paper concludes with a discussion on analyses, limitations, and future directions.

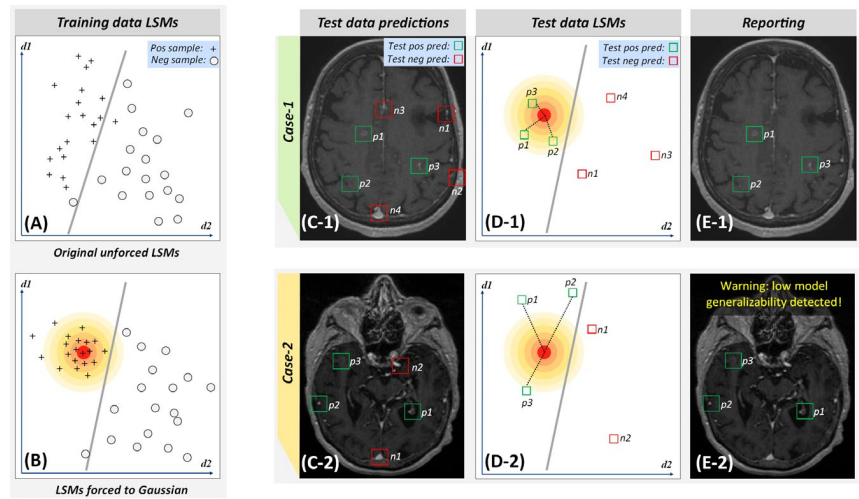

*Fig. 1. A simplified 2D representation of the latent space of multiple tumor candidates used to train a standard deep neural network; (A) in which positive (+) and negative (o) tumor candidates can be near-completely differentiated by a hyperplane that separates these, where the distribution does not follow a known statistical form, and (B) in which (+) candidates are mapped to a multivariate normal distribution. (C-E) During the inference, if the majority of the (+) candidates are not outliers (p1-p3 on image D-1), the model predicts that it generalizes for the case: Case 1 (C-1 through E1). If the majority of the (+) candidates are outliers with regards to the forced-Normal distribution (p1-p3 on image D-2), the model predicts that it has low generalizability for the case: Case-2 (C-2 through E2). In the event of low generalizability, the system displays a warning message to emphasize its uncertainty (E-2).*

## 2. Materials and methods

2.1 BM detection framework overview

A framework was introduced in [37] with the main goal of detecting smaller BM (<15mm) in 3D T1c datasets, which remains a challenging task due to the tumors' (1) smaller dimensions, (2) low contrast with surrounding tissues, and (3) visual similarities with vascular structures in some slice angles [38]. It consists of two stages: (1) candidate selection and (2) classification of the candidates as BM or not. The candidate selection stage adapts the scale-space point detection approach from [39] by integrating a minimax optimization to maximize the BM detection sensitivity while minimizing false-positive tumor detections. The classification stage employs a dedicated classification neural network called CropNet to differentiate BM from other candidates in 3D. Due to the under-representation of BM among the set of potential candidates, the framework uses a paired training strategy, where pairs of cubic regions centered at randomly selected positive and negative samples (i.e., BM and candidates that are not BM) form the training data batches. The training batches are then augmented on the fly (via random translations, rotations, gamma corrections, and elastic deformations [40]), and binary cross-entropy loss was minimized iteratively to optimize the network's classification performance (see Fig. 2).

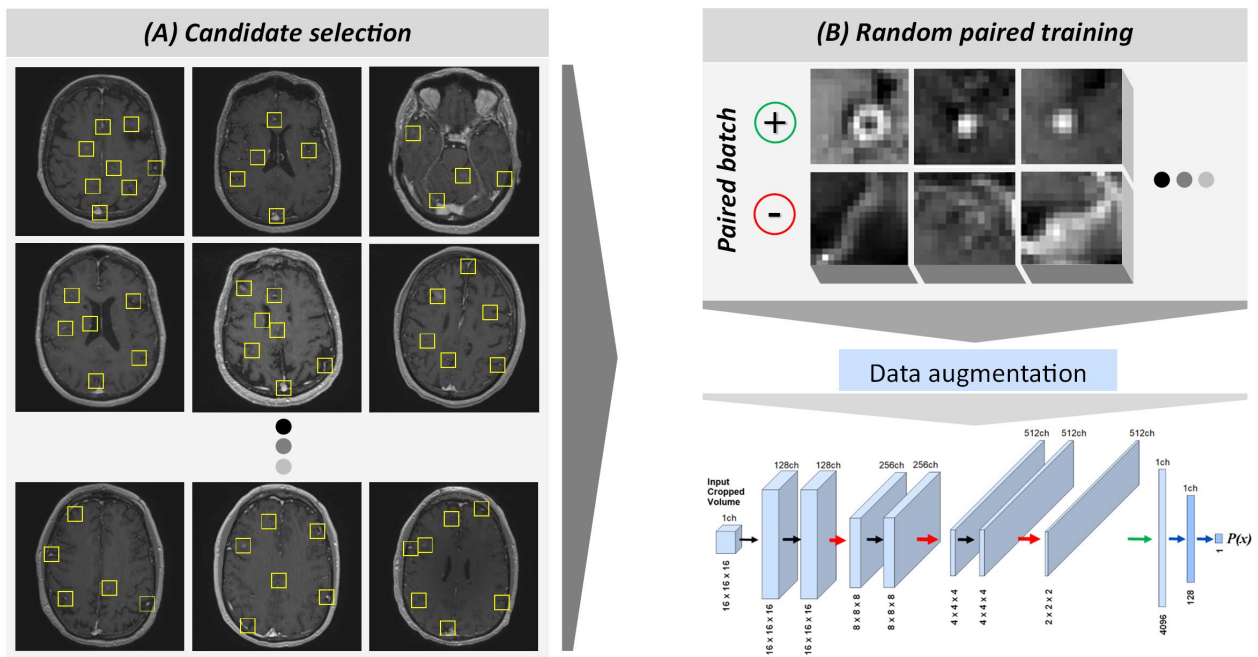

*Fig. 2. Training of the BM framework. (A) Candidate BM positions are computed using a scale-space-based point detection methodology in 3D; (B) cubic regions centered at the positive (+) (i.e., BM) and negative (-) (i.e., not BM) candidate positions are compiled into paired batches, augmented, and iteratively fed into CropNet optimized using binary cross-entropy loss.*

2.2. Technical contribution: Forced latent space mappings

The generalizability of an AI model may be conceptualized by specifying the underlying data representation held by the model. Domain shift, a common cause of reduced generalizability [26], [41], describes a mismatch between the underlying probabilistic distribution functions (PDFs) of training and new (i.e., test, unseen) data. Understanding the training data's PDF by observing the model's latent space is not an intuitive task, as the latent space mappings (LSMs) of modern DNNs are commonly not formulated to convey a specific distribution pattern. We hypothesize that if the data could be forced into a predefined PDF held by the model during its training, then the unseen data PDF divergence from this distribution can be quantified, leading to a metric for predicting the model's generalizability on the fly.

For a training dataset $x$ and a corresponding output $y$ (with unknown PDFs of $P(x)$ and $P(y)$ respectively)), a trained classifier network (i.e., CropNet) (1) maps $x$ into hidden layer outputs (i.e., LSMs) of $z_1, z_2, \cdots z_d$ with $d$ giving the network depth, and (2) produces the network output $\tilde{y}$ approximating $y$. The hidden layer PDFs are given by posterior distributions of $p(z_1|x,y), p(z_2|x,y), \cdots p(z_d|x,y)$, and the output PDF is given by $p(\tilde{y})$. Each $p(z_i|x,y)_{i \in [\![1,d]\!]}$ describes the model's representation of the underlying training data after accumulating all the data transformations specified in layer $i$ and all preceding layers; however, the PDFs of latter layers are more relevant as they represent the information distilled towards the target output. Accordingly, we refer to $p(z_d|x,y)$ as the underlying training data presentation, and LSMs as the latent space mappings of layer $d$. As mentioned previously, CropNet performs a binary classification task with the classes of 1:BM and 0:No-BM, where the positively labeled part of the training data could be shown as $(x_+, y_+)$. This study aims to force the underlying representation of only the positive part of the training data, as the BM class is heavily underrepresented in this specific application; the candidate selection stage generates ~60K candidates for a given 3D dataset where only a minuscule amount of them are BM centers [42]. To force the underlying data presentation of the positive part into a standard multivariate normal distribution (i.e., with zero mean and identity covariance matrix) as

$$p(z_{d+}|x_+, y_+) \approx N(0, I), \tag{1}$$

we introduce a Fréchet Normal Loss (FNL) that enables the given approximation during the model's training iteratively. For a given batch of LSMs from positive samples ($\bar{z} \subseteq z_{d+}$), FNL (1) computes the batch mean ($\mu_b$) and covariance matrix ($\Sigma_b$) of $\bar{z}$ and (2) returns the Fréchet distance [43] ($d^2$) between the batch's distribution and $N(0, I)$ as;

$$d^2 = |\mu_b|^2 + tr(\Sigma_b + I - 2 \cdot (\Sigma_b + \varepsilon \cdot I)^{1/2}), \tag{2}$$

with $\varepsilon$ denoting a small floating number to ensure that the square root of the term could be computed. Hence, it penalizes proportionally to the posterior distribution's divergence from $N(0, I)$. After the training, $p(z_{d+}|x_+, y_+)$ is estimated via a multivariate normal distribution $N(\mu_f, \Sigma_f)$ using the LSMs' final positions (i.e., mean and covariance across all positive entry mappings) without a reduction in dimension count.

For a given unseen data (e.g., a 3D MRI exam), the underlying training data representation of the predicted positive samples (i.e., pseudo positives) are analyzed: The Mahalanobis distances between LSMs coming from pseudo positive samples $\tilde{z}$ (with the corresponding network output of $\tilde{y} > \psi$, where $\psi$ is a network threshold calibrated for a specific BM detection sensitivity based on the training data) and the forced distribution (i.e., $N(\mu_f, \Sigma_f)$) are computed to give a set $D$ of Mahalanobis distances of predicted positives for a given MR exam to the forced distribution. Finally, the hypothesis that the majority of $D$ are not outliers with regards to the forced distribution is tested by comparing the median of $D$ versus a high quantile (i.e., 95%) chi-square distribution with degrees of freedom given by $dim(\mu_f)$. If the majority of $\tilde{z}$ are outliers, it indicates that the given unseen data may differ from the training data in its underlying presentation characteristics and result in poor model generalization for a given MR exam.

## 2.3. Database

The study database originated from two sources (1) the data acquired internally from the authors' institution and (2) Stanford University School of Medicine Brain Mets Dataset [36]. The study selection criteria were that the study (1) should include at least one BM and (2) should not include a BM with a diameter greater than 15mm. The internal dataset was collected retrospectively following Institutional Review Board approval with a waiver of informed consent (institutional IRB ID: 2016H0084). It consisted of 217 post-gadolinium T1-weighted 3D MRI (T1c) exams (contrast agent: gadoterate meglumine - 0.1 mmol/kg of body weight) collected from 158 patients with their BM segmentation masks. The Stanford dataset was collected from [36] and filtered (1) by the aforementioned study selection criteria and (2) to include only the T1 gradient-echo post images; giving 72 T1c exams. The data is divided into three groups for the analyses:

- *Training:* 175 T1c exams from 127 randomly selected patients of the internal dataset (75% of the internal dataset patients): 89 patients with one, 29 patients with two, 8 patients with three and a patient with four exams. The group's mean BM diameter is 5.6 mm (σ = 2.8 mm), median BM diameter is 4.7 mm, mean BM volume is 177 mm$^3$ (σ = 291 mm$^3$), and median BM volume is 55 mm$^3$.
- *Test-Int:* 42 T1c exams from 31 patients of the internal dataset (25% of the internal dataset patients) without patient overlap with the training set: 24 patients with one, 4 patients with two, 2 patients with three and a patient with four exams. The group's mean BM diameter is 4.6 mm (σ = 1.9 mm), median BM diameter is 4.1 mm, mean BM volume is 81 mm$^3$ (σ = 166 mm$^3$), and median BM volume is 36 mm$^3$.
- *Test-Stanford:* The complete set of the Stanford T1c exams. The group's mean BM diameter is 5.1 mm (σ = 2.2 mm), median BM diameter is 4.5 mm, mean BM volume is 121 mm$^3$ (σ = 224 mm$^3$), and median BM volume is 49 mm$^3$.

The histograms for the (1) BM count per exam, (2) BM diameter, and (3) BM volume for these groups are provided in Fig. 3. The spatial distributions of BM for the groups are presented in Fig. 4; the visualization was performed by adopting the strategy from [4] to present BM probability distributions on a template T1c exam.

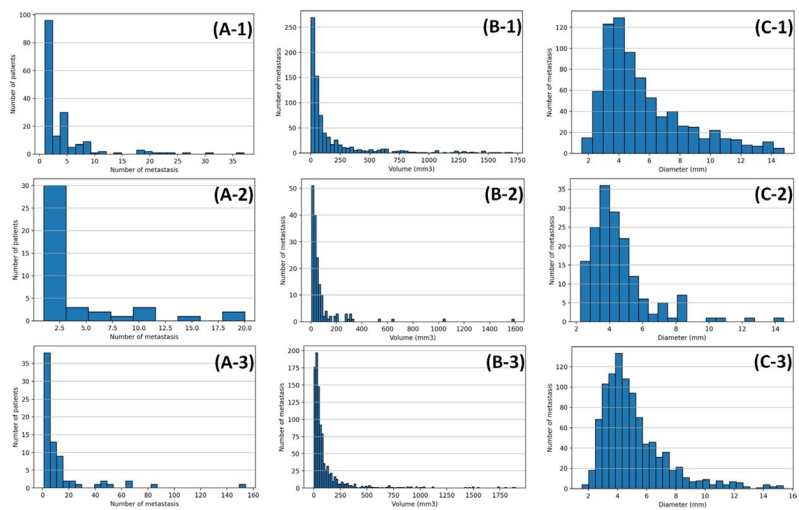

*Fig. 3. The histograms for the BM count per exam (A), BM volume (B) and BM diameter (C) are shown for Train (ABC-1), Test-Int (ABC-2) and Test-Stanford (ABC-3) groups.*

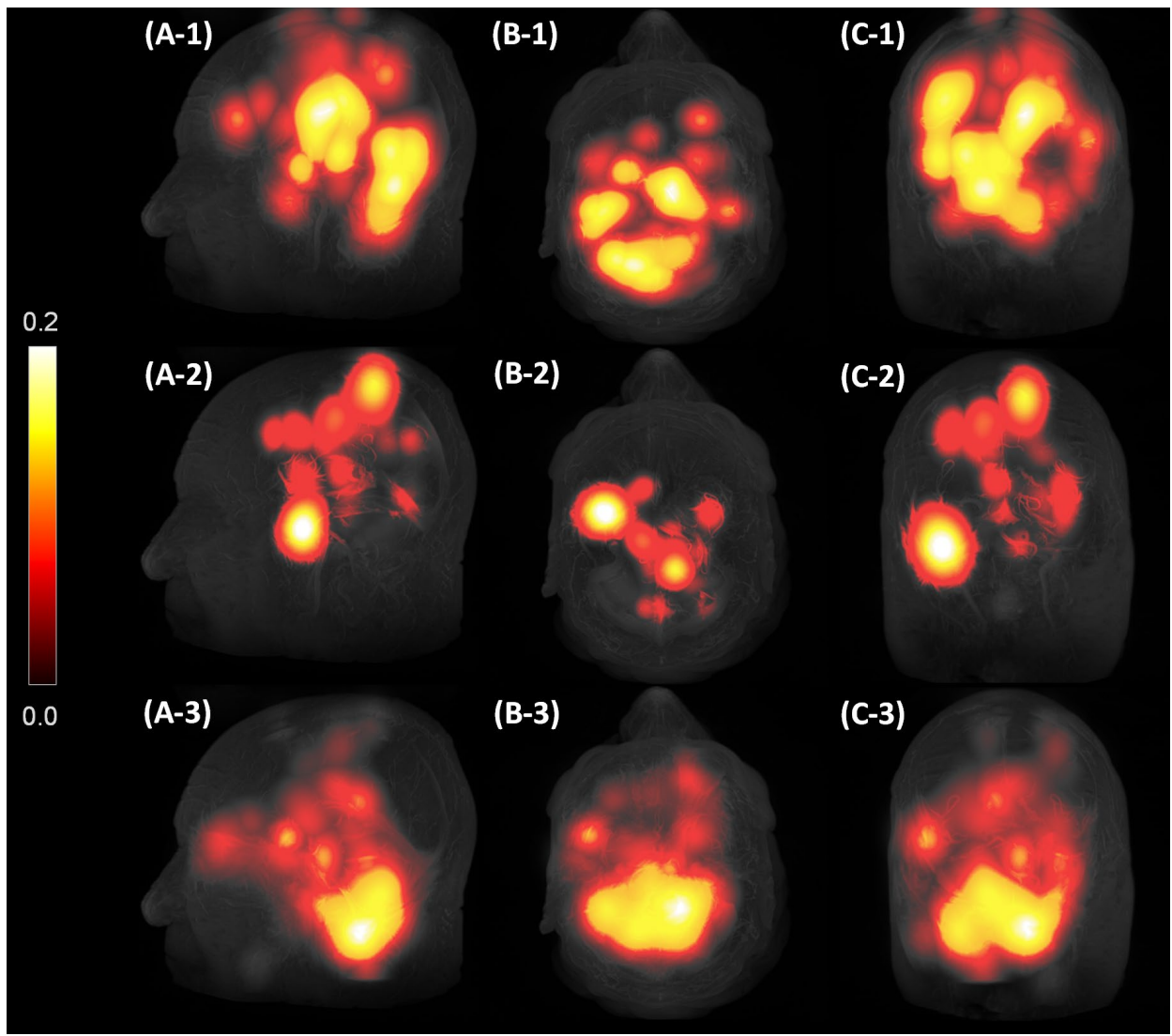

*Fig. 4. BM probability density function's projections on left sagittal (A), axial (B), and coronal (C) planes are provided for the Train (ABC-1), Test-Int(ABC-2) and Test-Stanford (ABC-3) groups.*

## 2.4. Validation metric

The average number of false positives-AFP (i.e., mean false BM detection count per exam) in connection with the detection sensitivity (i.e., the percentage of true BMs detected) was used as the validation metric. A tumor was marked as detected if the distance between a framework-generated detection and the tumor center was <1.5 mm. This metric provides a relevant measurement for the algorithm's applicability in real-life deployment scenarios as (1) the sensitivity of a detection system is critical, and (2) the number of false positives needs to be minimized to ensure the system's feasibility. Thus, various BM detection studies (including [44]–[46]) have utilized this metric.

## 3. Results

3.1 Validation study

The BM detection framework was trained using the training group. The candidate selection process generated ~72K BM candidates per exam capturing ~95% of the actual BM centers. The BM classification network, CropNet [4], was modified to produce four-dimensional LSMs as its secondary output (in addition to the BM probability output), as described in Section 2.2. The training process optimized the combination of (1) binary cross-entropy loss computed on the BM probability and (2) FNL computed on the LSM outputs. The weights for the loss components were 0.9 and 0.1, respectively, and determined empirically. The optimization was performed using the Adam algorithm [47], where the learning rate was 0.00005, and the exponential decay rates for the first and second moment estimates were set at 0.9 and 0.999. The training batch size was set at 128. After the network's training, a threshold value ψ was set to yield BM detection sensitivity of 90% for the training data.

Next, the model was executed on the testing groups (i.e., Test-Int and Test-Stanford). For each test exam, (1) the model produced the predicted BM locations and (2) the model's predicted generalizability status (i.e., high or low model generalizability), computed by processing the LSMs of a given exam as previously described. The association between AFP and BM detection sensitivity for the subgroups of predicted high and low model generalizability are reported in Fig. 5 and Table 1. The model predicted its generalizability as low for (1) 2 out of 42 Test-Int exams (4.8%) and (2) 33 out of 72 Test-Stanford exams (48.5%). For all testing data (i.e., the combination of Test-Int and Test Stanford), it produced (1) ~13.5 false positives (FPs) at 76.1% BM detection sensitivity for the low and (2) ~10.5 FPs at 89.2% BM detection sensitivity for the high generalizability groups, respectively.

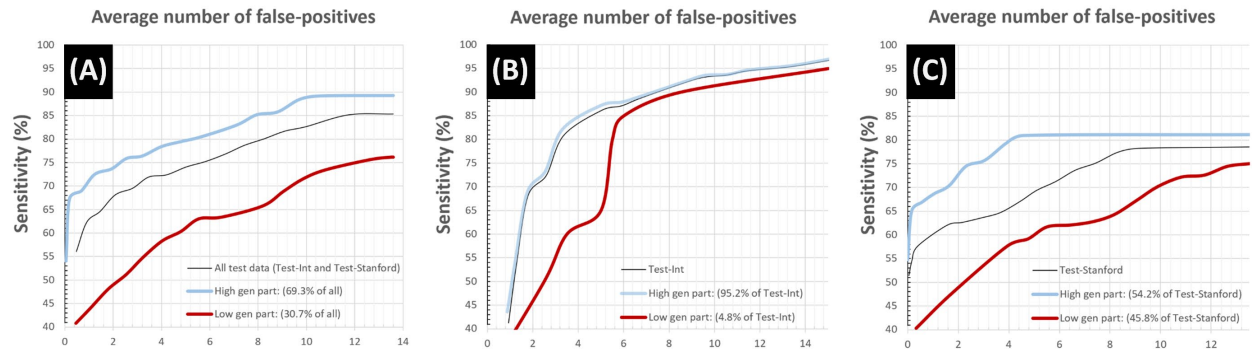

Fig. 5. AFP vs sensitivity for (A) the combination of two test groups, (B) Test-Int and (C) Test-Stanford. (1) The black curves: the complete set of exams; (2) the blue curves: the subgroup of exams for which the model predicted high generalizability; (3) the red curves: the subgroup of exams for which the model predicted its low generalizability.

3.2 Visualizations of LSMs

The decision curve (estimated based on the network's BM probability output) and LSMs of the actual BMs for the training, Test-Int and Test-Stanford groups are presented in Fig. 6. For ease of visualization, the figures only show the first two dimensions of the LSMs rather than all four dimensions. The BM LSMs' average $L^2$ norm, median distance to origin, and Fréchet distance to multivariate Normal distribution (i.e., computed via FNL) were (1) 1.201, 1.146, and 0.322 for the training, (2) 1.321, 1.288, and 0.432 for Test-Int, and (3) 2.682, 1.686, and 1.896 for Test-Stanford, respectively. In Fig. 7, the LSMs for a set of sample inputs from the Test-Int set are demonstrated; the mid-axial slice of the candidate cubic regions are presented in connection with the first two latent space coordinates.

TABLE 1: AFP VS SENSITIVITY

| Sensitivity %<br>Group | 70% | 75% | 80% | 85% | 87.5% | 90% |
|---|---|---|---|---|---|---|
| All test data | 2.93 | 5.66 | 8.21 | 11.63 | NA | NA |
| All test data – Low gen | 9.4 | 12.13 | NA | NA | NA | NA |
| All test data – High gen | 0.84 | 2.31 | 5.24 | 7.87 | 9.39 | NA |
| Test-Int | 2.14 | 2.83 | 3.28 | 4.67 | 6.12 | 7.51 |
| Test-Int – Low gen | 5.17 | 5.33 | 5.5 | 6 | 7.25 | 8.5 |
| Test-Int – High gen | 1.97 | 2.69 | 3.12 | 4.25 | 5.39 | 7.29 |
| Test-Stanford | 5.36 | 7.38 | NA | NA | NA | NA |
| Test-Stanford – Low gen | 9.84 | NA | NA | NA | NA | NA |
| Test-Stanford – High gen | 1.51 | 2.63 | 4.17 | NA | NA | NA |

*AFP counts at different sensitivity values are reported for all test data and Test-Int & Test-Stanford parts of the test data. 'High gen' refers to the subgroup of the dataset for which the model predicted high generalizability, and 'Low gen' refers to the subset for which the model predicted its low generalizability.*

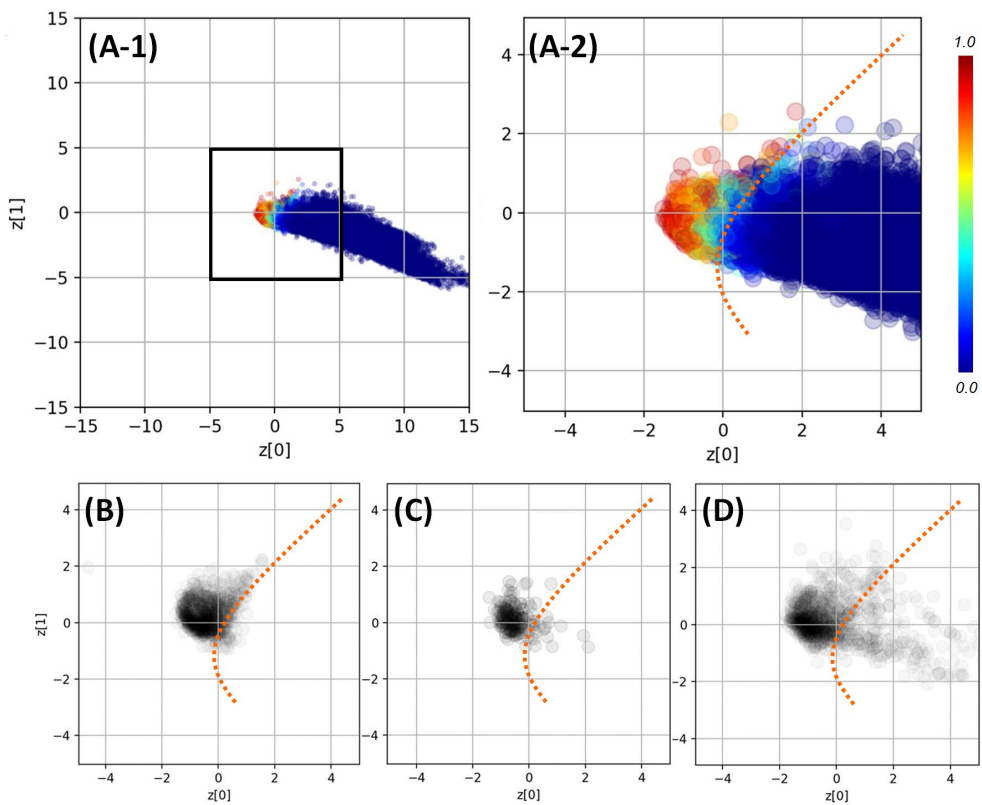

*Fig. 6. (A-1) The network's BM probability output is represented as a heat map for random candidate regions in latent space, (A-2) the decision curve is predicted based on the probability outputs, shown with an orange dashed curve. (B) Training, (C) Test-Int, and (D) Test-Stanford BM LSMs are shown with the predicted decision curve.*

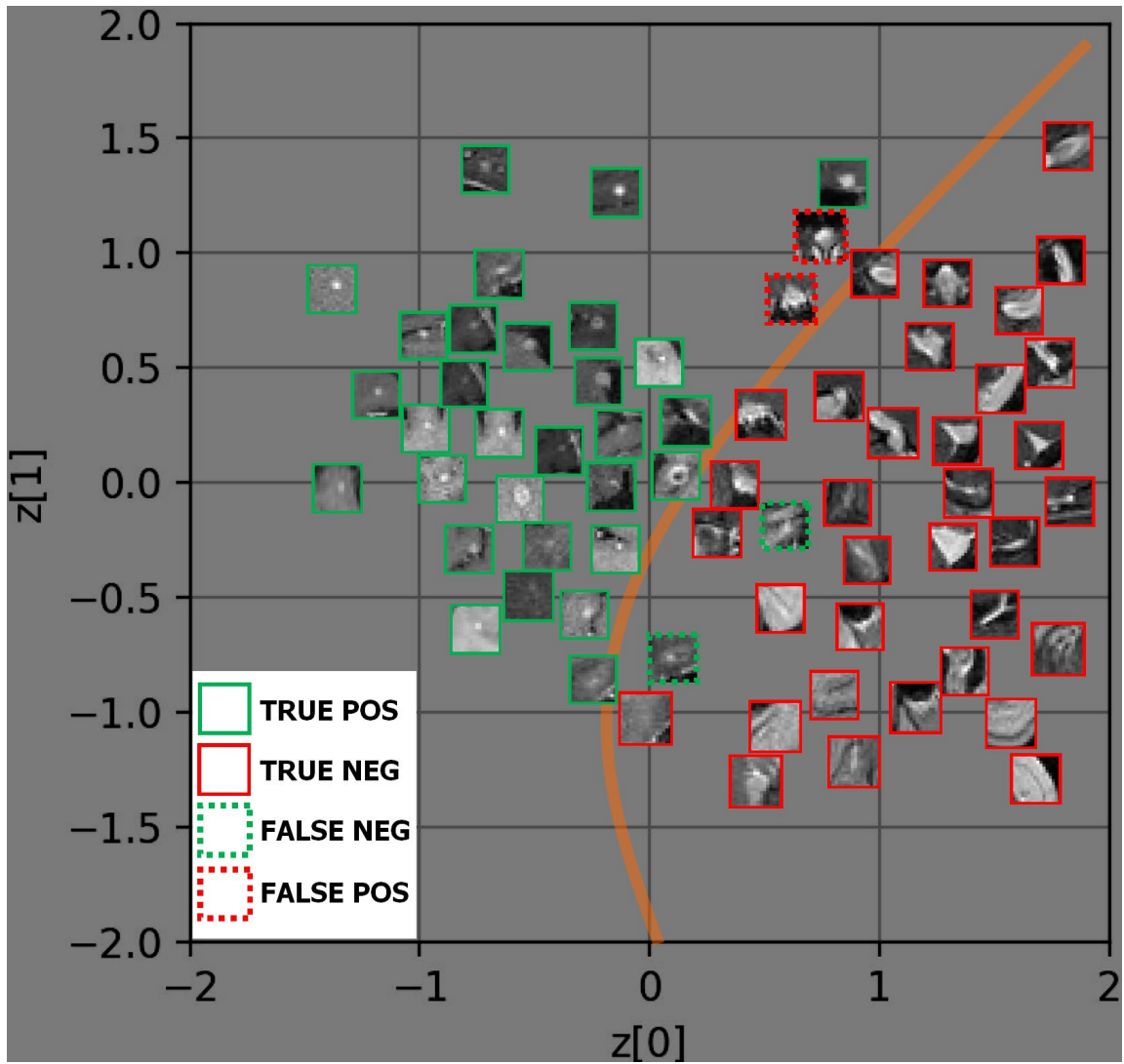

*Fig. 7. Mid-axial slices of a sample set of candidate cubic regions from the Test-Int group are presented at coordinate positions specified by the first two dimensions of their LSMs (only 2 of the 4 latent space dimensions are used in this visualization for simplicity). True positives: solid green, true negatives: solid red, false positives: dashed red, and false negatives: dashed green. The decision curve is presented in orange.*

### 3.3 Radiology workflow integration

The integration of medical AI solutions into radiology workflows was examined in various studies [48]–[50]. In [50], three maturity levels were proposed for radiology workflows that integrate AI; research, production, and feedback. These levels reflect the readiness and infrastructure of an institution as described in further detail in the referenced paper. We briefly describe and adapt the research workflow to present a sample integration of the introduced approach.

At a common Radiology workflow, imaging modalities (e.g., MRI, CT) send acquired images to a DICOM router distributing received images to pertinent storage locations, such as the PACS or VNA. To benefit from the proposed AI-based algorithm, the radiologist may send images to a DICOM node where the BM detection framework is deployed. In our deployment, the DICOM node is a virtual machine running an application composed of a DICOM listener and Python script implementation of the framework. The framework receives the input DICOM images (corresponding to the T1c dataset), processes them, and prepares the results as two Grayscale Softcopy Presentation State (GSPS) objects [51]:

(1) *BM detection result GSPS:* combining the graphic data (i.e., circles) to present the detected BM centers in their corresponding DICOM image positions, and textual data to show detected BMs' identifiers with the algorithm ID.
(2) *Model generalization status GSPS:* textual data to present the model's generalization status for the given exam rendered at the upper left side of the exam's slices (see Fig. 8).

After the GSPS objects are generated, they are sent to a separate Research-PACS; thus, the official image records in PACS, as well as patient EMRs, remain intact. The archives in the Research-PACS are accessible by stand-alone advanced DICOM viewers that allow viewing and analyzing BM detection results in connection with their corresponding standard DICOM images.

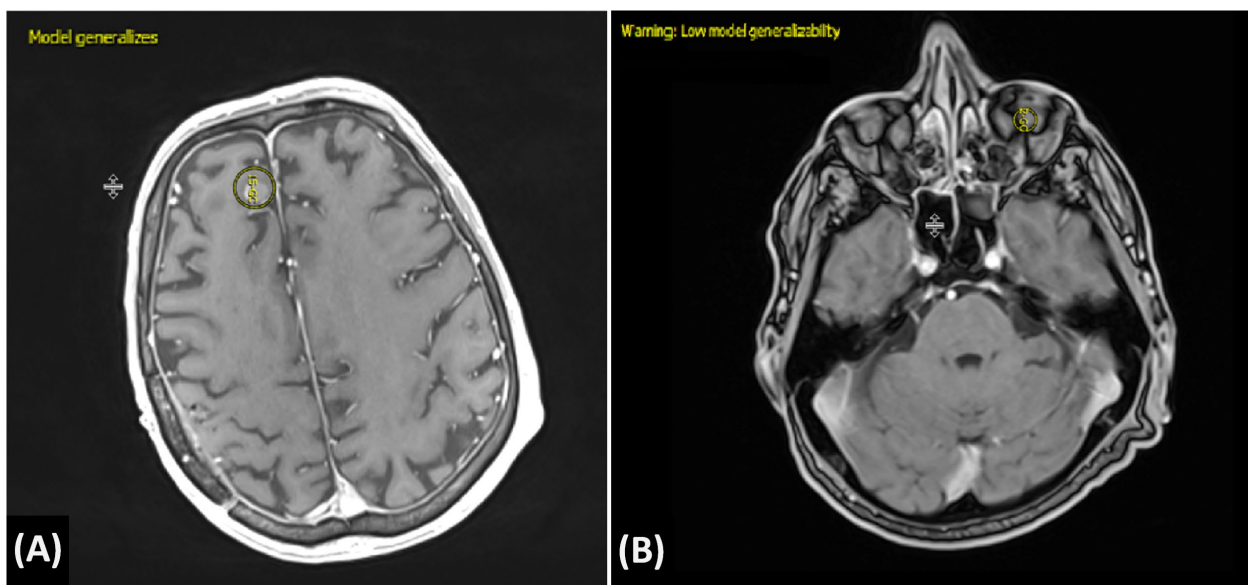

Fig. 7. The locations of the BM and the model's generalization status are overlaid on two different exams by an advanced DICOM viewer; (A) 'model generalizes' text is generated by the proposed algorithm for exams for which it predicts high generalizability, (B) A 'low model generalizability' warning is generated by the proposed algorithm in cases of low generalizability. In this case, the MRI exam was from another organization acquired with a scanner different from that the scanners used in the training data.

## 4. Discussion and conclusion

The introduced approach enabled detection of reduced AI model generalizability when applied to new cases. As presented in Fig. 5 and Table 1, the model had reduced performance for the MRI exams for which the model detected low model generalizability. The lower generalizability partition of the Test-Stanford group produced 9.84 false positives at 70% sensitivity (failing to achieve 75% detection sensitivity), whereas the detected high generalizability partition of the same group produced 4.17 false positives at 80% detection sensitivity (see Table 1, Test-Stanford rows). For the Test-Int group, results similar to the Test-Stanford group were reported (see Table-2, Test-Int rows). Only two exams in the Test-Int group were classified as low model generalizability (as the AI model was trained with the internally acquired data). Future studies may (1) examine AI model generalizability and performance for T1c BM datasets collected from additional institutions and (2) examine the causes of generalizability discrepancies, such as scanner parameters and patient cohort characteristics.

To generate Figs. 6-B, C, and D., (1) cubic regions centered at actual BM locations in the Training, Test-Int and Test-Stanford were fed into the trained network, and (2) the first two dimensions of the resulting LSMs were shown. Fig 6-B shows that most of the cubic regions known to be BMs based on ground-truth annotations were on the left side of the decision curve (or on the corresponding side of the surface in 4D) determined by the training data. Based on the dimensions shown, it resembles a normal 2D distribution visually, producing FNL of 0.322. The LSMs of BM in Test-Int (see Fig. 6-C) also showed an approximately normal distribution (FNL: 0.432), whereas the LSMs of BM in Test-Stanford (see Fig. 6-D) presented a distribution significantly differing from the training data (FNL: 1.896). These findings may suggest an alternative utilization for the proposed method; to predict the generalizability of an AI model (trained using the introduced method) for a novel dataset (e.g., collected from a different institution, acquired with a different scanner or scanner parameters, etc.), a labeled subset (i.e., annotated with known positives) of this dataset could be processed to generate an FNL value. A higher FNL value would indicate that the AI model poorly maps the BM data in the latent space; hence, the model would likely not generalize well to the novel dataset, regardless of its accuracy for the given subset. This additional feature could be further investigated in future studies.

We presented the applicability of the proposed technical innovation by integrating it into a two-stage detector by (1) forcing the candidate volumetric areas' LSMs into a specific form during its training and (2) processing test candidate LSMs to determine the model's generalizability on the fly. However, this approach could also be employed in an end-to-end solution, which uses a neural network taking the whole 3D volume as an input and producing segmentation/detection results as an output (such as DeepMedic [52] utilized for BM segmentation/detection in multi-sequence MRI in [44] and [53]). To this end, the selected network's last fully connected layer output could be regulated using the introduced methodology. Accordingly, a given input volume will be represented by a single point in the latent space, where the point's Mahalanobis distance to the forced mapping distribution's origin (i.e., the mean) could be used as a metric to determine the model's generalizability for the given input. The integration of the proposed method into an end-to-end solution may be examined in later studies.

In this paper, we introduced an approach to enable an AI model to predict its own generalizability for previously unseen data. Accordingly, we described (1) a forced LSM technique utilizing a novel Fréchet Normal Loss for the model training and (2) the processing of LSMs to determine the model's generalizability. The merit of this approach was presented using a BM detection framework with the T1c data collected from both internal and external sources. In addition to the technical innovation introduced, this study focused on the increasingly important field of model generalizability, where the specific concept of estimation of model generalizability after the model's deployment was investigated, with promising results.

# References

[1] I. El Naqa, M. A. Haider, M. L. Giger, and R. K. Ten Haken, "Artificial Intelligence: reshaping the practice of radiological sciences in the 21st century," *Br. J. Radiol.*, vol. 93, no. 1106, p. 20190855, 2020.

[2] S. K. Zhou *et al.*, "A review of deep learning in medical imaging: Imaging traits, technology trends, case studies with progress highlights, and future promises," *Proc. IEEE*, vol. 109, no. 5, pp. 820–838, 2021.

[3] S. Jang *et al.*, "Deep learning--based automatic detection algorithm for reducing overlooked lung cancers on chest radiographs," *Radiology*, vol. 296, no. 3, pp. 652–661, 2020.

[4] E. Dikici *et al.*, "Automated Brain Metastases Detection Framework for T1-Weighted Contrast-Enhanced 3D MRI," *IEEE J. Biomed. Heal. Informatics*, p. 1, 2020.

[5] E. J. Hwang *et al.*, "Development and validation of a deep learning--based automatic detection algorithm for active pulmonary tuberculosis on chest radiographs," *Clin. Infect. Dis.*, vol. 69, no. 5, pp. 739–747, 2019.

[6] X. Liu, K. Chen, T. Wu, D. Weidman, F. Lure, and J. Li, "Use of multimodality imaging and artificial intelligence for diagnosis and prognosis of early stages of Alzheimer's disease," *Transl. Res.*, vol. 194, pp. 56–67, 2018.

[7] G. Muscogiuri *et al.*, "Artificial intelligence in coronary computed tomography angiography: from anatomy to prognosis," *Biomed Res. Int.*, vol. 2020, 2020.

[8] S. Gupta and Y. Kumar, "Cancer prognosis using artificial intelligence-based techniques," *SN Comput. Sci.*, vol. 3, no. 1, pp. 1–8, 2022.

[9] N. L. Eun *et al.*, "Texture analysis with 3.0-T MRI for association of response to neoadjuvant chemotherapy in breast cancer," *Radiology*, vol. 294, no. 1, pp. 31–41, 2020.

[10] D. Russo *et al.*, "Prediction of chemo-response for serous ovarian cancer using DNA methylation patterns with deep machine learning (AI)," *Gynecol. Oncol.*, vol. 162, p. S240, 2021.

[11] C. Li *et al.*, "Deep learning-based AI model for signet-ring cell carcinoma diagnosis and chemotherapy response prediction in gastric cancer," *Med. Phys.*, vol. 49, no. 3, pp. 1535–1546, 2022.

[12] F. Maleki, K. Ovens, R. Gupta, C. Reinhold, A. Spatz, and R. Forghani, "Generalizability of Machine Learning Models: Quantitative Evaluation of Three Methodological Pitfalls," *arXiv Prepr. arXiv2202.01337*, 2022.

[13] H. Salehinejad *et al.*, "A real-world demonstration of machine learning generalizability in the detection of intracranial hemorrhage on head computerized tomography," *Sci. Rep.*, vol. 11, no. 1, pp. 1–11, 2021.

[14] V. M. T. de Jong, K. G. M. Moons, M. J. C. Eijkemans, R. D. Riley, and T. P. A. Debray, "Developing more generalizable prediction models from pooled studies and large clustered data sets," *Stat. Med.*, vol. 40, no. 15, pp. 3533–3559, 2021.

[15] A. C. Justice, K. E. Covinsky, and J. A. Berlin, "Assessing the generalizability of prognostic

information," *Ann. Intern. Med.*, vol. 130, no. 6, pp. 515–524, 1999.

[16] T. Eche, L. H. Schwartz, F.-Z. Mokrane, and L. Dercle, "Toward Generalizability in the Deployment of Artificial Intelligence in Radiology: Role of Computation Stress Testing to Overcome Underspecification," *Radiol. Artif. Intell.*, vol. 3, no. 6, p. e210097, 2021.

[17] J. Futoma, M. Simons, T. Panch, F. Doshi-Velez, and L. A. Celi, "The myth of generalisability in clinical research and machine learning in health care," *Lancet Digit. Heal.*, vol. 2, no. 9, pp. e489--e492, 2020.

[18] B. Neyshabur, S. Bhojanapalli, D. McAllester, and N. Srebro, "Exploring generalization in deep learning," *Adv. Neural Inf. Process. Syst.*, vol. 30, 2017.

[19] D. Anderson and K. Burnham, "Model selection and multi-model inference," *Second. NY Springer-Verlag*, vol. 63, no. 2020, p. 10, 2004.

[20] A. D'Amour *et al.*, "Underspecification presents challenges for credibility in modern machine learning," *arXiv Prepr. arXiv2011.03395*, 2020.

[21] S. Mutasa, S. Sun, and R. Ha, "Understanding artificial intelligence based radiology studies: What is overfitting?," *Clin. Imaging*, vol. 65, pp. 96–99, 2020.

[22] C. Shorten and T. M. Khoshgoftaar, "A survey on image data augmentation for deep learning," *J. big data*, vol. 6, no. 1, pp. 1–48, 2019.

[23] F. Zhuang *et al.*, "A comprehensive survey on transfer learning," *Proc. IEEE*, vol. 109, no. 1, pp. 43–76, 2020.

[24] R. Moradi, R. Berangi, and B. Minaei, "A survey of regularization strategies for deep models," *Artif. Intell. Rev.*, vol. 53, no. 6, pp. 3947–3986, 2020.

[25] E. W. Steyerberg and F. E. Harrell, "Prediction models need appropriate internal, internal--external, and external validation," *J. Clin. Epidemiol.*, vol. 69, pp. 245–247, 2016.

[26] E. Kondrateva, M. Pominova, E. Popova, M. Sharaev, A. Bernstein, and E. Burnaev, "Domain shift in computer vision models for MRI data analysis: an overview," in *Thirteenth International Conference on Machine Vision*, 2021, vol. 11605, pp. 126–133.

[27] H.-J. Yoo, "Deep convolution neural networks in computer vision: a review," *IEIE Trans. Smart Process. Comput.*, vol. 4, no. 1, pp. 35–43, 2015.

[28] V. Sze, Y.-H. Chen, T.-J. Yang, and J. S. Emer, "Efficient processing of deep neural networks: A tutorial and survey," *Proc. IEEE*, vol. 105, no. 12, pp. 2295–2329, 2017.

[29] L. Oakden-Rayner, "Exploring Large-scale Public Medical Image Datasets," *Acad. Radiol.*, vol. 27, no. 1, pp. 106–112, 2019.

[30] N. L. S. T. R. Team, "The national lung screening trial: overview and study design," *Radiology*, vol. 258, no. 1, pp. 243–253, 2011.

[31] A. E. Flanders *et al.*, "Construction of a machine learning dataset through collaboration: the RSNA 2019 brain CT hemorrhage challenge," *Radiol. Artif. Intell.*, vol. 2, no. 3, 2020.

[32] P. Dluhos *et al.*, "Multi-center Machine Learning in Imaging Psychiatry: A Meta-Model Approach," *Neuroimage*, vol. 155, 2017.

[33]   K. Simonyan and A. Zisserman, "Very deep convolutional networks for large-scale image recognition," *arXiv Prepr. arXiv1409.1556*, 2014.

[34]   K. He, X. Zhang, S. Ren, and J. Sun, "Deep residual learning for image recognition," in *Proceedings of the IEEE conference on computer vision and pattern recognition*, 2016, pp. 770–778.

[35]   M. J. Willemink *et al.*, "Preparing medical imaging data for machine learning," *Radiology*, vol. 295, no. 1, p. 4, 2020.

[36]   C. for Artificial Intelligence in Medicine & Imaging, "BrainMetShare, available online at: https://aimi.stanford.edu/brainmetshare, last accessed on 11.05.2021." .

[37]   E. Dikici, M. Bigelow, R. D. White, B. S. Erdal, and L. M. Prevedello, "Constrained generative adversarial network ensembles for sharable synthetic medical images," *J. Med. Imaging*, vol. 8, no. 2, p. 24004, 2021.

[38]   E. Tong, K. L. McCullagh, and M. Iv, "Advanced imaging of brain metastases: from augmenting visualization and improving diagnosis to evaluating treatment response," *Front. Neurol.*, vol. 11, p. 270, 2020.

[39]   T. Lindeberg, "Scale selection properties of generalized scale-space interest point detectors," *J. Math. Imaging Vis.*, vol. 46, no. 2, pp. 177–210, 2013.

[40]   P. Y. Simard, D. Steinkraus, and J. C. Platt, "Best practices for convolutional neural networks applied to visual document analysis," in *Seventh International Conference on Document Analysis and Recognition, 2003. Proceedings.*, 2003, vol. 1, pp. 958–963.

[41]   E. H. P. Pooch, P. L. Ballester, and R. C. Barros, "Can we trust deep learning models diagnosis? The impact of domain shift in chest radiograph classification," *arXiv Prepr. arXiv1909.01940*, 2019.

[42]   E. Dikici, X. V Nguyen, M. Bigelow, and L. M. Prevedello, "Augmented networks for faster brain metastases detection in T1-weighted contrast-enhanced 3D MRI," *Comput. Med. Imaging Graph.*, vol. 98, p. 102059, 2022.

[43]   D. C. Dowson and B. V Landau, "The Frechet distance between multivariate normal distributions," *J. Multivar. Anal.*, vol. 12, no. 3, pp. 450–455, 1982.

[44]   O. Charron, A. Lallement, D. Jarnet, V. Noblet, J.-B. Clavier, and P. Meyer, "Automatic detection and segmentation of brain metastases on multimodal MR images with a deep convolutional neural network," *Comput. Biol. Med.*, vol. 95, 2018.

[45]   E. Grovik, D. Yi, M. Iv, E. Tong, D. Rubin, and G. Zaharchuk, "Deep learning enables automatic detection and segmentation of brain metastases on multisequence MRI," *J. Magn. Reson. Imaging*, vol. 51, 2019.

[46]   Z. Zhou *et al.*, "Computer-aided detection of brain metastases in T1-weighted MRI for stereotactic radiosurgery using deep learning single-shot detectors," *Radiology*, vol. 295, no. 2, pp. 407–415, 2020.

[47]   D. Kingma and J. Ba, "Adam: A Method for Stochastic Optimization," *Int. Conf. Learn. Represent.*, 2014.

[48]   E. Ranschaert, L. Topff, and O. Pianykh, "Optimization of Radiology Workflow with Artificial Intelligence," *Radiol. Clin.*, vol. 59, no. 6, pp. 955–966, 2021.

[49]  J. H. Sohn *et al.*, "An open-source, vender agnostic hardware and software pipeline for integration of artificial intelligence in radiology workflow," *J. Digit. Imaging*, vol. 33, no. 4, pp. 1041–1046, 2020.

[50]  E. Dikici, M. Bigelow, L. M. Prevedello, R. D. White, and B. S. Erdal, "Integrating AI into radiology workflow: levels of research, production, and feedback maturity," *J. Med. Imaging*, vol. 7, no. 1, p. 16502, 2020.

[51]  National Electrical Manufacturers Association (NEMA), "Digital Imaging and Communications in Medicine (DICOM)—Supplement 33: Grayscale Softcopy Presentation State (GSPS) Storage," Rosslyn, VA, 1999.

[52]  K. Kamnitsas *et al.*, "Efficient multi-scale 3D CNN with fully connected CRF for accurate brain lesion segmentation," *Med. Image Anal.*, vol. 36, pp. 61–78, 2017.

[53]  Y. Liu *et al.*, "A deep convolutional neural network-based automatic delineation strategy for multiple brain metastases stereotactic radiosurgery," *PLoS One*, vol. 12, no. 10, p. e0185844, 2017.

[54]  E. Dikici, X. V Nguyen, M. Bigelow, J. L. Ryu, and L. M. Prevedello, "Advancing Brain Metastases Detection in T1-Weighted Contrast-Enhanced 3D MRI Using Noisy Student-Based Training," *Diagnostics*, vol. 12, no. 8, p. 2023, 2022.